\documentclass[%
 reprint,
 amsmath,amssymb,
 aps,
]{revtex4-1}

\usepackage{graphicx}
\usepackage{dcolumn}
\usepackage{bm}
\usepackage{color}

\begin{document}
\preprint{APS/123-QED}

\title{Causality, Uncertainty Principle, and Quantum Spacetime Manifold in Planck Scale}

\author{Hamidreza Simchi}
\email{simchi@alumni.iust.ac.ir}
\affiliation {Department of Physics, Iran University of Science and Technology, Narmak, Tehran 16844, Iran} \affiliation{Semiconductor Technoloy Center, P.O.Box 19575-199, Tehran, Iran}

\date{\today}

\begin{abstract}
In causal set theory, there are three ambiguous concepts that this article tries to provide a solution to resolve these ambiguities. These three ambiguities in Planck's scale are: the causal relationship between events, the position of the uncertainty principle, and the kinematic. Assuming the interaction between events, a new definition of the causal relationship is presented. Using the principle of superpoisition, more than one world line are attributed to two events that are interacting with each other to cover the uncertainty principle. Using these achievements, it is shown that kinematics has no place in the Planck dimension and that quantum spacetime manifold should be used instead.
\\
\\
Keywords: Causal set theory, Causality, Uncertainty Principle, Kinematic, Manifold 
\end{abstract}

\maketitle


\section{Introduction}
Scientists are pursuing three main areas of researches to achieve the theory of quantum gravity. The oldest is the theory of quantum strings, which due to the impossibility of evaluating its results in the laboratory, advances in this field are considered more from the perspective of mathematics{[}1{]}. The second theory is Loop Quantum Gravity, which does not seem to be a comprehensive theory at Planck scale{[}1{]}. The theory of causal sets is the third important branch of researches that has been considered by scientists today due to the use of simple and fundamental assumptions{[}2{]}. The important point for the complete success of these theories in the field of quantum gravity seems to be the concept of time in physics which remains unresolved up to now. In the other words, one of the main unsolved problem of physic is the true nature of time. Some scientists believe that all that exists are things that change. Things do not change in time; the change of things is time and time is simply a complex of rules that govern the change. Time is inferred from things {[}3{]}. Others believe that everything that is true and real is such in a moment that is one of a succession of moments. Space is emergent and approximate and the laws of nature evolve in time and may be explained by their history. Time is the most real aspect of our perception of the world {[}4{]}.  In our previous article, we have concluded that {[}5{]}:

\begin{enumerate}
\item  The world is composed by events that change. 

\item  We sense the changes of events as the passage of time.

\item  All events which are in mutual or multi-interaction with each other compose a system and other non-related events compose its environment. A boundary exists between each system and its environment.  

\item  In each application domain of a physical theory, there are some main conceptual paradigms. During the transition between the different application domains through the boundaries, one should pay enough attention to the conceptual paradigm shift.
\end{enumerate}

It should be noted that before formulating the theory of causal sets in the form that is now available to us, important and fundamental researches have been done by scientists. Robb has defined null, parallel lines and plane and prove numerous theorems involving them and described the relativity using the discrete spacetime (i.e., casual structure) {[}6,7{]}. Hawking et al., {[}8{]} and Malament {[}9{]} have proved that the casual structure of a spacetime, together with a conformal factor, determine the metric of a Lorentzian spacetime, uniquely. It has been shown that one can recover the conformal metric by using the before and after relations amongst all events {[}10{]}. Now, if one has a measure for the conformal factor, he/she can recover the entire metric and spacetime {[}10{]}. Ofcourse, ${}^{,}$t Hooft {[}11{]} and Myrheim {[}12{]} have independently found the causal set theory too.
\\
Of course, other efforts are being made by scientists to introduce the theory of quantum gravity by attention to the locality and causality. One of them is causal dynamic triangulation (CDT){[}13{]}. Near the Planck scale, the structure of spacetime itself is supposed to be constantly changing due to quantum fluctuations and topological fluctuations. CDT theory uses a triangulation process which varies dynamically and follows deterministic rules, to map out how this can evolve into dimensional spaces similar to that of our universe{[}13{]}. Diel{[}14{]} has assumed that the elementary structure of spacetime is a derivative of causal dynamical triangulation and, at the elementary level, space consists of a (discrete) number of interconnected space points, each of which is connected to a small number of neighbouring space points. He has shown that emergence and propagation of quantum fields (including particles) are mapped to the emergence and propagation of space changes by utilizing identical paths of in/out space point connections{[}14{]}. Also, it is well known that in Einstein's theory of general relativity, events are placed on the system world line, and Schrodinger's time-dependent equation emphasizes the existence of a causal relationship between events. On the other hand, based on the particle approach in quantum mechanics, as well as describing the quantum field theory and many body physics by particle creation/annihilation operators of particles/quasi-particles, the issue of locality can be considered as an important subject. In other words, by considering the causality and locality, it is possible to develop an alternative causal model of quantum theory and quantum field theory, in which quantum objects are the basic units of causality and locality{[}15{]}. In this model, not only the quantum objects are embedded in space and move within space, but also the dynamics of space is triggered by the dynamics of the quantum objects. The causal model of QT/QFT assumes discretized spacetime similar to the spacetime of causal dynamical triangulation{[}15{]}.
\\
In this paper, we try to answer three ambiguous concepts in the theory of causal sets at Planck scale. These three problems are determining the type of causal relationship between events, explaining the position of the uncertainty principle and its importance in quantizing the theory of causal sets and the place of kinematics in this theory. First, with a brief review of the theory of causal sets, we enumerate the basic features of this theory. Then, with a brief review of the concept of causality in physics, we explain the type of causal relationship between events in the theory of causal sets to use in the rest of this article. By reviewing the effect of the constant speed of light on the theory of special relativity and its relation to the concept of time in physics, we show that kinematics can have no place in the Planck scale. Finally, considering the position of the uncertainty principle in quantum physics and reviewing published articles in the field of quantum manifold, we will compile and introduce the general structure of the quantum spacetime manifold.
\\
The structure of the article is as follows: in section II, we review the discrete spacetime as casual sets. A short review about the special causality in physics is presented in section III and in section IV the kinematical and dynamical models are discussed. The property of quantum spacetime manifold is provided in section V and the summary is presented in section VI.

\section{Discrete spacetime as causal set}

\noindent \noindent In a causal set $C$ including the elements $\{a_1,a_2,a_3,\cdots ,a_{n-1},a_n\}$ the relation $a_i<a_j$ for $i\le j$ is satisfied. The pair $(C,\le )$ is reflexive, antisymmetric, transitive, and locally finite.  Therefore, the causal matrix $C$ can be defined by
\begin{equation} \label{GrindEQ__1_} 
C_{a_i,a_j}=\{ \begin{array}{c}
1,\ \ \ \ \ \ \ \ \ \ \ a_i<a_j \\ 
0\ \ \ \ \ \ Otherwise \end{array}
\end{equation} 
Also, a nearest neighbor relation (called link) is a relation $a_i<a_j$ such that there exists no $a_k\in C$ with $a_i<a_k<a_j$. The elements $a_i$ and $a_j$ are the nearest neighbors and their relation is shown as $a_i<*a_j$. Now, the link Matrix $L$ can be defined by
\begin{equation} \label{GrindEQ__2_} 
L_{a_i,a_j}=\{ \begin{array}{c}
1,\ \ \ \ \ \ \ \ \ \ \ a_i<*a_j \\ 
0\ \ \ \ \ \ Otherwise \end{array}
\end{equation} 
It is obvious that both $C$ and $L$ matrices are strictly upper triangular and a causal set is a partially ordered set. By attention to the relativistic causality [16,17], one can construct a causal set from a Lorentzian manifold $(M,g)$. The manifold $M$ represents the collection of all spacetime events and the metric $g$ is a symmetric non-degenerate tensor on $M$ of signature $(+,-,-,-)$. We know, the infinitesimal displacement is given by
\begin{equation} \label{GrindEQ__4_} 
{ds}^2=-{dt}^2+{\delta }_{ij}{dx}^i{dx}^j 
\end{equation} 
where, $i,j=1,2,3,\cdots ,d$ and here $d=1$. We can rewrite Eq. \eqref{GrindEQ__4_} as
\begin{equation} \label{GrindEQ__5_} 
{ds}^2=-\left(dt+dx\right)\left(dt-dx\right)+{\delta }_{ij}{dx}^i{dx}^j 
\end{equation} 
where, $i,j=1,2,3,\cdots ,d-1$. By defining, $x^+=\frac{(x+t)}{\sqrt{2}}$ and $x^-=\frac{(t-x)}{\sqrt{2}}$ , we can write
\begin{equation} \label{GrindEQ__6_} 
{ds}^2=-2{dx}^+{dx}^-+{\delta }_{ij}{dx}^i{dx}^j 
\end{equation} 
By comparing Eq. \eqref{GrindEQ__6_} with Eq. \eqref{GrindEQ__4_}, it can be concluded that both $x^+$ and $x^-$ act as time-coordinate. It is called the lightcone coordinate. One nice thing about the lightcone coordinate is that the causal structure is partially included into the coordinate system itself. Therefore, for two points $x_1=(x^+_1,x^-_1)$ and $x_2=(x^+_2,x^-_2)$ we have $x_1\le x_2$ if and only if $x^+_1\le x^+_2$ and $x^-_1\le x^-_2$. Now, If the length of diamond in lightcone coordinate be equal to $S$, one can find the $n$ random points in the (1+1) dimensional space by
\begin{widetext}
\begin{equation} \label{GrindEQ__7_} 
P=S\ \times \ Random\ number\left(x^-,x^+\right)\ \times \ Rotation\ matrix\ (45^\circ) 
\end{equation} 
\end{widetext}
It should be noted that
\begin{widetext}
\begin{equation} \label{GrindEQ__8_} 
\left( \begin{array}{c}
t \\ 
x \end{array}
\right)=\left( \begin{array}{cc}
\sqrt{2}/2 & \sqrt{2}/2 \\ 
-\sqrt{2}/2 & \sqrt{2}/2 \end{array}
\right)\left( \begin{array}{c}
x^- \\ 
x^+ \end{array}
\right)=\left( \begin{array}{cc}
\cos 45^\circ & \sin 45^\circ \\ 
-\sin 45^\circ & \cos 45^\circ \end{array}
\right)\left( \begin{array}{c}
x^- \\ 
x^+ \end{array}
\right)\  
\end{equation}
\end{widetext} 
For example, we found 1000 points in a (1+1)-dimensional space by  
\begin{widetext}
\begin{equation} \label{GrindEQ__9_} 
P=1\ \times \ Random\ number\left(-0.5,+0.5\right)\ \times \ Rotation\ matrix\ (45^\circ) 
\end{equation} 
\end{widetext}
 and shown them in Fig.1, after sorting. 

\begin{figure}[]
\includegraphics[width=\columnwidth]{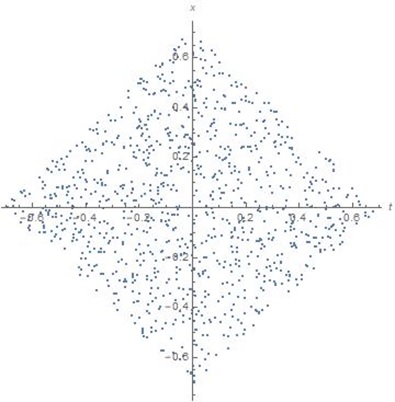}
\caption{\label{fig:epsart} (Color online) 1000 random points in (1+1)-dimensional space.}
\end{figure}

\noindent However, in (1+1)-dimensional there are one temporal (unidirectional) dimension and one spatial (bidirectional) dimension. Since, the proper time is given by
\begin{equation} \label{GrindEQ__10_} 
{d\tau }^2=-{dt}^2+dx^2_i 
\end{equation} 
For, $dt>0$ and ${d\tau }^2>0$, the points will be placed in future timelike region. It means that not only the spatial distance ($dx^2_i$) should be grater than the temporal distance (${dt}^2$) but also $dt>0$. Therefore, the element of the casual matrix $C$ will be equal to one if the both conditions are satisfied simultaneously for two elements $a_i$ and $a_j$ of the causal set and otherwise it will be equal to zero. Using the method, one can find the causal matrix $C$. By keeping the non-zero elements of $C$ -matrix when $a_i$ and $a_j$ are only the nearest neighbor elements and replacing the other non-zero elements by zero number, the link matrix $L$ can be found. The above explained method which is used for finding the causal set, the causal matrix and the link matrix from a Lorentzian manifold is called sprinkling method. 

\noindent Since, the points of a casual set are placed in the future timelike region, it can be concluded that there is a priority (time precedence) between points respect to the time of occurrence. In the other words, a finite path of length $n$ (maximum chain) is a sequence of distinct elements $a_1<*a_2<*a_3<*\cdots <*a_{n-1}<*a_n$ in the future timelike region. Therefore, the priority in occurrence is called the causality in casual set theory. The causal set which is found from a Lorentzian manifold by sprinkling method is invariant under the boost transformation in spite of the lattice model. Therefore, the causal set based physical theory is Lorentzian invariant at Planck scale in spite of the other physical theories about spacetime at Planck scale. In next section, we will discuss about the causality in physics and show that the priority in occurrence is the sufficient condition and the interaction between each two relates  $a_i<*a_j$ is the necessary condition for assigning the causal relation to two relates $a_i$ and $a_j$.\textbf{}

\section{Causality in physics}

\noindent
 In Newtonian physics, one can exactly determine the future if he/she knows the initial and boundary conditions. The process is called a deterministic process. Time-dependent Schrodinger equation is a deterministic equation i.e., if one knows the initial and boundary conditions at time $t$ he/she will be able to find the state function of the system at time $t+1$. But, in quantum physics, the total state of a system is specified by the superposition of substates (superposition principle). Based on the principle, nobody knows the exact final state of the system before observation. After observation, one of the superposed substates will create the output of observation. The process is called a probabilistic process. In probabilistic process the output of observation can be created by one of the many superposed substates and in deterministic process the output of observation is created by the exact initial state of the system. Therefore, there is an interaction process between output and input of observation such that the output is created by input while we cannot exactly specify the input before appearing the output in the probabilistic process. The interaction between output and input is called causality. In Newtonian physics, there is the deterministic causality and in quantum physics there is the probabilistic causality. Therefore, in deterministic causality, the elements of the casual world line have two properties: causality and priority in occurrence (time precedence). But in probabilistic causality, we encounter many world lines theoretically (before observation) such that the elements of each causal worldline have the causality and priority properties. It means that, a causal set which is found by sprinkling method and have a specific finite path has only the priority properties and cannot be considered as a deterministic causality. For classical point particle, we assign a specific path  $a_1<*a_2<*a_3<*\cdots <*a_{n-1}<*a_n$ to the system in the future timelike region. Therefore, the specific path in a causal set which is found by sprinkling method is not a suitable candidate for the probabilistic causality. For a quantum point particle, we should consider all chains between $a_1$ and $a_n$ and then use the discrete path integral method for finding the amplitude for the whole trajectory [13]. But we did not consider the causality between events in this case, and in consequence we lost some important information or added some non necessary information to the final state of the closed system including observer. It means that the current sprinkling method for arising the causal set is only suitable for the deterministic causality (classical systems) if the causal relation will be added to it.

\noindent {\color{red}{Also}}, causality is an interaction process between input and output although it has a certain concept between folks.  Usually, folks have some intuitions about causality. The raised question is: whether there is something in the world that realizes the intuition of folk about the causality? The question has to be answered empirically, and thus commonly depends on the natural science. It is called Canberra methodology [18]. The Canberra methodology includes two stages [18,19]. At first stage, we specify something which we interested to analyze them from philosophical point of view. Then we collect together the platitudes concerning our subject matter and finally conjoin them for defining a theoretical role for the things we are interested in. At the second stage, we look at our theory of the world to tell us what, if anything, plays the rule so defined [18,19]. Ofcourse, there is another methodology which is called naturalism [18]. The methodology is often divided into a descriptive and a normative part [20,21]. In the descriptive part it is studied how we acquire knowledge within science and in normative part the justification for this knowledge is given [18,20,21]. The naturalistic approach to causation has become well known as the empirical analysis of causation [18]. It has been shown that there is no difference between two methodologies about causation if we consider the causation as interaction between relates and pay attention to the fact that output of observation is created by its input [18]. Therefore, the elements of causality worldline have two important properties. First, there is an occurrence priority between them and second the prior relata causes the next relata. It means that we should omit the non-causal elements from the worldline for finding the causal worldline. The causal world line shows the history of system evolutions in the future timelike region. If we deal with the quantum physics, we have to consider all causal world lines between two relates before observation for showing the probabilistic history of system evolutions in the future timelike region due to the superposition principle. Ofcourse, from the Heisenberg uncertainty principle point of view, we have to consider more than a causal world line before observation, too. Therefore, for quantum point particle we should use the discrete path integral method for finding the amplitude for the whole trajectory [22]. Since, we consider the causal world line in our closed system including observer the time passes as changes in relates. But in kinematic model of causal set theory since we only consider the priority in occurrence theoretically the time does not pass because no changes happen in the relates. By attention to the new concept of time (as change in relates) we review the kinematic and dynamic models in the next section. 
   
\section{Kinematical or dynamical models}

\noindent

In physics, the kinematic is referred to the time independent case. If time is sensed as the change of things, kinematic will be equal to the no change case. In the other words, if no change is sensed no time will pass and in consequence defining the time is meaningless. It can be shown that the special relativity can be deduced from the assumption that the velocity of light does not depend on the observer and it is the maximum velocity of things in vacuum [23]. In order to make the concept of time clearer let us, assume two frame of references $A$ and $B$ move with velocity $v$ respect to each other. The observers on both references have no sense about time in own reference frame but when they see the other frame since its position changes, he/she sense the time. Also, let us, assume two rulers are placed in each frame. If they want to measure the length of ruler in own frame, they can use two light flashes. The time difference between received flashes from the back and the front of the ruler multiplied by the velocity of light $C$ in own frame is equal to the length of the ruler. It should be noted that, in the closed system including ruler, light flashes and observer the change in position of light flashes is sensed and therefore time passes. For measuring the length of moving ruler, they should measure the time difference between received flashes from the back and the front of the ruler, again. But, whether the rate in the change of the flash positions is equal to the previous case? i.e., whether the velocity of light $C$ does not depend on observer frame of reference? Why?

\noindent Let us, assume $C\ $is constant (note that it is only a assumption). Fig.2 shows the spacetime diagram of two moving reference frames respect to each other. At time $T$, the observer in nonmoving frame sends a light flash toward the moving frame. The observer in moving frame receives the flash at time $t$${}_{2}$. The light flash is reflected toward the nonmoving frame by a mirror and the observer receive it at time $k^2T$. The equation of moving of light flash (red arrow) is 
\begin{equation} \label{GrindEQ__11_} 
t-T=\frac{x}{C} \to x=C(t-T) 
\end{equation} 
and the equation of moving observer is
\begin{equation} \label{GrindEQ__12_} 
t=\frac{x}{v}\to x=vt 
\end{equation} 
In the triangle with two red arrows, the dashed blue line is the middle-perpendicular line  and in consequence one can write
\begin{equation} \label{GrindEQ__16_} 
\frac{t_1}{t_2}=\frac{C\sqrt{C-v}}{\left(C-v\right)\sqrt{C+v}}=\frac{1}{\sqrt{1-v^2/C^2}} 
\end{equation} 
It means that the assumption of independency of light velocity to reference frame causes the time dilation. Now, if the length of ruler in moving frame is  $L_0$ (the ruler is at the rest) its length in nonmoving frame (ruler is moving) can be calculated as
\begin{equation} \label{GrindEQ__18_} 
L=\frac{Cv}{C-v}\frac{L_0}{kv}=\frac{CL_0}{C-v}\frac{\sqrt{C-v}}{\sqrt{C+v}}=\frac{L_0}{\sqrt{1-v^2/C^2}} 
\end{equation} 

Therefore, the assumption of independency of light velocity to the reference frame causes the length contraction.
\begin{figure}[]
\includegraphics[width=\columnwidth]{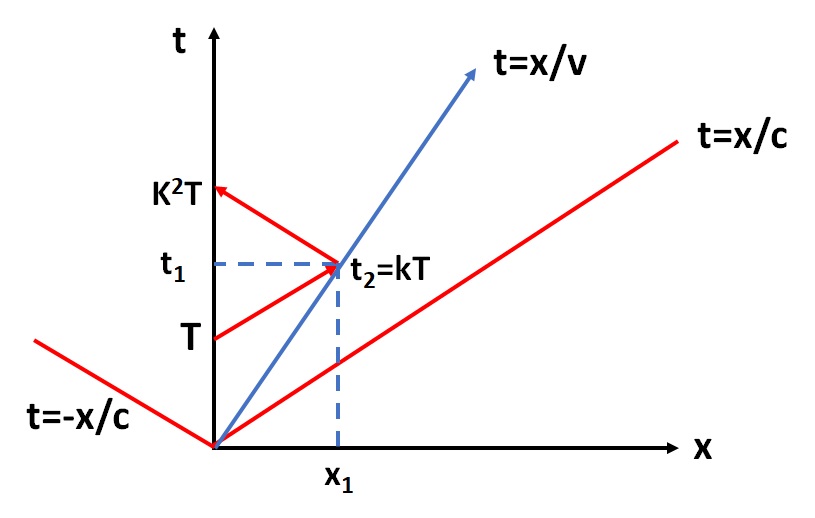}
\caption{\label{fig:epsart}  (Color online) The spacetime diagram of two moving reference frames respect to each other. Red arrows show the light flashes and the blue arrow shows the causal world line of moving frame.}
\end{figure}

But, in relativity, proper time $(\tau)$ along a timelike world line is defined as the time as measured by a clock following that line. It is thus independent of coordinates, and is a Lorentz scalar.The proper time interval between two events on a world line is the change in proper time. This interval is the quantity of interest, since proper time itself is fixed only up to an arbitrary additive constant, namely the setting of the clock at some event along the world line. The proper time interval between two events depends not only on the events but also the world line connecting them, and hence on the motion of the clock between the events. It is expressed as an integral over the world line (analogous to arc length in Euclidean space). An accelerated clock will measure a smaller elapsed time between two events than that measured by a non-accelerated (inertial) clock between the same two events. The twin paradox is an example of this effect.
\\
Therefore, up to now, we used two main assumptions and one definition: 

\begin{enumerate}
\item  If nothing changes in a closed system, the time definition is meaningless. It is called the dynamical assumption.

\item  If the velocity of light is constant and maximum velocity of things in vacuum, we expect to see time dilation and length contraction phenomena. It is called the velocity of light assumption [23].

\item The proper time interval between two events depends not only on the events but also the world line connecting them. 
\end{enumerate}

\noindent Then, from special relativity point of view the below questions can be asked:

\begin{enumerate}
\item  What is about the dynamical assumptions at the Planck scale? 

\item  Whether it is correct that the causal set dynamic is found from a kinematic version of a causal set if the kinematic version, which includes no time, cannot exist at the Planck scale? 

\item  What is about the velocity of light assumption at the Planck scale? 

\item  Whether it is expected that we see some physical phenomena related to the non-variable velocity of light at the Planck scale? 
\end{enumerate}

Although it is not possible to answer all of these questions by using the above explanations, but even if we consider the proper time, since kinematics means time independency and quantum gravity theory is supposed to explain spacetime on the Planck scale, this theory cannot be based on a kinematical theory and should be developed based on a dynamical theory from the begining.

\section{Quantum manifold of spacetime}

\noindent It has been shown that two very different manifold could not approximate the causal set, and in general, an arbitrary causal set may not embed in any Lorentzian manifold with a metric [24]. The question about how manifoldlike causal sets may arise from suitable dynamical laws has been justified, before [25,26]. Generally, there are three types of dynamics that a causal set can has [26]. The classical dynamic can be used for explaining the continuum limit which is the general relativity. The dynamics of quantum matter and fields on a given ``classical'' causal set can be used for explaining the continuum limit which is the quantum field theory on a fixed curved spacetime. Finally, quantum dynamics of the causal set itself, which is the final aim in order to construct a quantum theory for gravity [26]. But, is there a kinematical discrete spacetime at Plank scale such that the both general relativity and quantum theory can be deduced from the spacetime? If one of the main aims of finding the quantum gravity theory is solving the existence of singularities in general relativity and renormalization requirement in quantum physics, why should one develop the classical dynamic and dynamic of quantum matter? It seems that the quantum dynamics of the causal set itself should be the main branch of the future research program. In this research program, we should find a quantum spacetime manifold for deducing a suitable discrete causal set when the time is defined based on the changes in the elements of the causal set.

\noindent

In above, we showed that for developing a causal set theory for quantum gravity at Planck scale, we should specify the importance and effectiveness of the below natural facts in our theory when we want to study the continuum limit:

\begin{enumerate}
\item  The maximum velocity of things in vacuum which is the velocity of light.

\item   Kinematic has no place in quantum gravity theory at Planck sacle.

\item  The uncertainty principle and superposition principle of quantum mechanics.
\end{enumerate}

\noindent In the other words, the new quantum spacetime manifold should has some special properties for providing the above three requirements at least at continuum limits. 

\noindent We know that the manifold geometry ($M$) is the heart of the general relativity and the observable operators on Hilbert space (Schwartz space $(S\left(R^n\right))$) are the main components of quantum mechanics. Since, $R^n$ is the space of the position of classical events, it is expected that the background space $R^n$ will be the limit of the $M$ and $S(R^n)$. Now let us, assume that there is an infinite quantum manifold $M_Q$. It is well known that the expectation values of quantum observable operators follow the classical laws. Therefore, it may be possible one recovers the manifold geometry $M$ from $M_Q$ by calculating the position expectation value [27,28]. Also, in parallel, $M_Q$ can be locally homomorphic to the $S(R^n)$ [27,28]. But, the square-integrability is very important in quantum physics and in consequence we should only consider the family of all functions which have the below property 
\begin{equation} \label{GrindEQ__19_} 
{\left|\left|f\right|\right|}_{\alpha ,\beta }={sup}_{x\in R^n}\left|x^{\alpha }D_{\beta }f(x)\right| 
\end{equation} 
For all multiindices $\alpha $ and $\beta $, it is a family of seminorms which generates a topology on $S(R^n)$. This topology is called the natural topology [27,28]. Now, if we define the position expectation value as $\overline{Q}=\frac{\left\langle f,\ Qf\right\rangle }{\left\langle f,f\right\rangle }$ , the open sets of expectation value topology (${\overline{Q}}^{-1}\left(W\right)$) exist and id defined as  
\begin{equation} \label{GrindEQ__20_} 
{\overline{Q}}^{-1}\left(W\right)=\left\{f\in S^{\neq 0}\, |\, \overline{Q}(f)\in W\right\}
\end{equation} 
where, $W\subset R^n$ is open in the standard topology on $R^n$. Thus, the expectation value topology is the coarsest topology in which the function $\overline{Q}$ is continuous [27,28]. By attention to the above definitions, it can be shown that the final quantum manifold will be a differentiable infinite dimensional manifold locally homeomorphic to $S^{\neq 0}(R^n)$ and in contrast to the usual definition of an atlas, two different topologies called expectation value topology and natural topology should be introduced [27,28]. Fig. 3 shows a quantum atlas, schematically.

\begin{figure}[]
\includegraphics[width=\columnwidth]{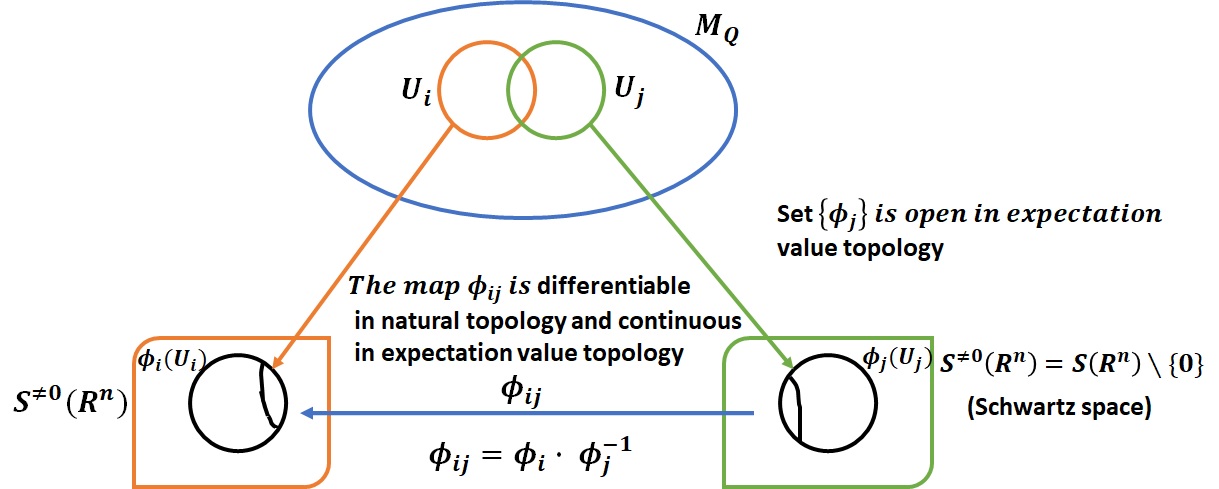}
\caption{\label{fig:epsart}   (Color online) The Schematic of a quantum atlas.}
\end{figure}

\noindent Now, a quantum manifold of dimension $n$ is a set $M_Q$ equipped with an equivalence class of quantum atlases of dimension $n$. The element of $M_Q$ are called quantum points [27,28]. If one find a suitable method for arising the causal set from the quantum manifold, he/she will have a quantum causal set as the fundamental network of a spacetime at Planck scale. Ofcourse, it can be a research program in future.

\section{Summary}

\noindent
We have encountered some important problems with physics which three of them seems to be the most important. The singularities in general relativity, the renormalization requirements in quantum physics and the concept of time. Some bodies believe that if we can solve the problem of time, the other two remained problems will be solved. However, we have discussed about the nature of time in our previous article (Ref.5) and concluded that the time can be sensed as the changes in things. It means that under kinematic condition time can not be defined, basically. Since, we are searching a unified theory between gravity and quantum for solving the above three mentioned main problems, at least, it seems that developing the dynamic of a causal set theory based on a kinematic causal set cannot help us much in this direction although, for studying some related classical problems at continuum level, it may help us. In the other words, we need a dynamical causal set at beginning. It means that a causal set should be raised from a quantum manifold. The quantum manifold is locally homomorphic to the Schwartz space and in parallel, the necessary manifold geometry of relativity can be recovered by using the quantum manifold. It should be noted that the causality relation differs from time precedence. In causality, two relates interact with each other and make change in each other but in time precedence, the priority is only important. Therefore, in a closed system including observer, we should consider a quantum manifold such that the causal world line, which is created by causal events between relates, appears in the manifold geometry of relativity. Also, we should pay enough attention to uncertainty and superposition principles for assigning a set of causal chains (paths) to each event instead of a specific exact path. Therefore, in Schwartz space, we should consider a superposition of square integrable functions with different amplitudes when we want to study the homomorphic condition. 

\section*{Acknowledgement}
We would like to thank Dr. Yasaman K. Yazdi for providing us the related code of Fig.1 and its related explanations.

\nocite{*}

\bibliography{apssamp}
 \textbf{References}

\noindent [1] Lee Smolin, ``The Trouble With Physics: The Rise of String Theory, The Fall of a Science, and What Comes Next'' (	Houghton Mifflin Harcourt, 2006).

\noindent [2] Sumati Surya, ``The causal set approach to quantum gravity'',  Springer \textbf{22}, 5 (2019).

\noindent [3] Julian Barbour, ``The End of Time'' (Oxford university press, 1999).

\noindent [4] Lee Smolin, ``Time Reborn'' (Houghton Mifflin Harcourt publishing company, 2013).

\noindent [5] Hamidreza Simchi, ``The concept of Time: A Grand Unified Reaction Platform'', arXiv:1910.07875v1 [physics.gen-ph] 27 Sep 2019, Philsci-archive.pitt.edu/18632/.

\noindent [6] A. A. Robb, `'A Theory of Time and Space'' (Cambridge University Press, 1914).

\noindent [7] A. A. Robb, `'Geometry of Time and Space'' (At the University Press, 1936).

\noindent [8] S. W. Hawking, A. R. King, and P. J. Mccarthy, ``A New Topology for Curved Space-Time which incorporates the causal, differential, and conformal structures'', J. Math. Phys. \textbf{17}, 174 (1976).

\noindent [9] D. B. Malament, `` The class of continuous timelike curves determines the topology of spacetime'', J. Mathem. Phys. \textbf{18}, 1399 (1977). 

\noindent [10] L. Bombelli, J. Lee, D. Meyer, and R. Sorkin, ``Space-Time as a Causal Set'', Phys. Rev. Lett. \textbf{59}, 521 (1987).

\noindent [11] G. ${}^{,}$t Hooft, ``Quantum Gravity: A Fundamental Problem and Some Radical Ideas'', Springer US, Boston 323 (1979).

\noindent [12] J. Myrheim, ``Statistical Geometry'', CERN-TH-2538.

\noindent [13] Jeff Murugan, Amanda Weltman, and George F. R. Ellis, ``Foundations of Space and Time'', (Cambridge University Press, 2012). 

\noindent [14] Hans H. Diel, ``A Local Causal Model of Spacetime Dynamics'', Open Access Library Journal \textbf{5}, e4957 (2018).

\noindent [15] Hans H. Diel, ``Spacetime Structures in a Causal Model of Quantum Theory'', Open Access Library Journal \textbf{4}, e3357 (2017).

\noindent [16]  R. Penrose, ``Technique of Differential Topology in Relativity'', SIAM (1972).

\noindent [17] S. W. Hawking and G. F. R. Ellis, ``The large-scale structure of space-time'' (Cambridge University Press, 1973).

\noindent [18] Jakob Sprinkerhof, ``Causal Structure in Quantum Field Theory'', (PhD Thesis, The University of Leeds, 2015).

\noindent [19] D. Nolan, ``Platitudes and metaphysics'' (MIT Press, 2009).

\noindent [20] W. V. Quine, ``Epistemology naturalized. In Ontological Relativity and Other Essays'' (Columbia University Press, New York, 1969).

\noindent [21] M. Devitt, ``Realism and Truth'' (Princeton University of Press, 1984).

\noindent [22] Steven Paul Johnston, ``Quantum Fields on Causal Sets'' (PhD Thesis, Imperial College London, 2010).

\noindent [23] Christoph Schiller, ``Motion Mountain: the adventure of physics-Vol.II, Relativity and Cosmology'' (Edition 31,  www.motionmountain.net, 2020).

\noindent  [24] R. D. Sorkin, ``Causal sets: Discrete gravity'', gr-qc/0309009

\noindent [25] J. Henson, ``The Causal set approach to quantum gravity'', gr-qc/0601121

\noindent [26] P. Wallden, ``Causal sets dynamics: Review and outlook'', J. Phys: Confere. Seri. \textbf{453}, 012023 (2013).

\noindent  [27] M. Hohmann, R. Punzi, and M. N.R. Wohlfarth, ``Quantum manifolds with classical limit'', arXiv:0809.3111v1 [math-ph] (2008).

\noindent [28] Manuel Hohmann, ``Quantum Manifold'', Foundation and Probability in Physics - 6 AIP Conf. Proc. \textbf{1424}, 149-153 (2012). 

\noindent 

\noindent

\end{document}